\documentclass[12pt,letter]{article}
\usepackage[section]{placeins}
\usepackage[table]{xcolor}
\usepackage{lipsum}
\usepackage{multirow}
\usepackage{setspace}
\usepackage{caption}
 \usepackage[square,numbers]{natbib}
\usepackage{anysize}
\marginsize{1.25in}{1.25in}{1in}{1in}
\usepackage{authblk}
\usepackage{graphicx}
\usepackage{color}
\usepackage{url}
\usepackage{latexsym}
\usepackage{amsmath}
\usepackage{amssymb}
\usepackage{amsfonts}
\usepackage{multirow}
\usepackage{bbm}
\usepackage{syntonly}
\usepackage{titlesec}
\usepackage{epstopdf}
\usepackage{graphics}
\usepackage{epsfig}
\usepackage{bm}
\usepackage{booktabs,multirow,tabularx}

\definecolor{brcolor}{rgb}{0.8,0.0,0.4}

\newcommand{\be}{\begin{eqnarray}}

\newcommand{\ee}{\end{eqnarray}}           \newcommand{\ba}{\begin{eqnarray*}}
\newcommand{\ea}{\end{eqnarray*}}

\titleformat{\section}
    {\centering\normalfont\Large\bfseries}{\thesection.}{1em}{}

\begin{document}
\title{\textbf{Non-decimated 2D Wavelet Spectrum and Its Use in Breast Cancer Diagnostics}}

\author[1]{Minkyoung Kang}
\author[2]{William Auffermann}
\author[3]{Brani Vidakovic\footnote{Corresponding Author: brani@stat.tamu.edu}}
\affil[1]{{\small \emph{Microsoft Research, 1 Microsoft Way, Redmond, WA 98052, USA}}}
\affil[2]{{\small \emph{Department of Radiology \& Imaging Sciences, University of Utah Health,  30 North 1900 East,   Salt Lake City, UT 84132}}}
\affil[3]{{\small \emph{Department of Statistics, Texas A\&M University, College Station, TX 77843-3143}}}
\date{}
\maketitle


\begin{abstract}
To improve diagnostic accuracy of breast cancer detection, several researchers have used the wavelet-based  tools, which provide additional insight and information for aiding diagnostic decisions. The accuracy of such diagnoses, however, can be improved. This paper introduces a wavelet-based technique, non-decimated wavelet transform (NDWT)-based scaling estimation, that improves scaling parameter estimation over the traditional methods. One distinctive feature of NDWT is that it does not decimate wavelet coefficients at multiscale levels resulting in redundant outputs which are used to lower the variance of scaling estimators. Another interesting feature of the
proposed methodology is the freedom of dyadic constraints for inputs, typical for standard wavelet-based approaches.

To compare the estimation performance of the NDWT method to a conventional orthogonal wavelet transform-based method, we use simulation to estimate the Hurst exponent in two-dimensional fractional Brownian fields. The results of the simulation show that the proposed method improves the conventional estimators of scaling and yields estimators with smaller mean-squared errors. We apply the NDWT method to classification of mammograms as cancer or control and,
for publicly available mammogram images from the database at the University of South Florida, find the the diagnostic accuracy in excess of 80\%.
\end{abstract}

\section{Introduction}
Breast cancer is the most common form of cancer in terms of incidence and the second most common form of cancer with regards to cancer mortality in women in the United States. The early detection of the breast cancer is crucial for patients' survival because the survival rates significantly improve with early detection and treatment (American Cancer Society, Cancer Facts \& Figures 2021). Mammographic screening is the most common means of breast cancer screening for the early detection of breast cancer. However, even experienced radiologists may not identify up to 30\% of mammograms because breast tissue is complex and signatures of disease can be subtle (Martin et al., 1979). In addition, multiple mammographic screenings might be required to confirm the cancer and each screening is costly and stressful to the subjects. A number of existing computer methods for breast cancer detection focus on the detection of specific markers, such as tumors or micro-calcifications. In this paper, we characterize the self-similar properties of normal breast tissue with NDWT, where non-normal tissue is a potential marker of breast cancer.

To assess the degree of self-similarity in a breast tissue area, we develop an estimation method based on non-decimated wavelet spectra. The parameters describing spectral regularity form a battery of spectral summaries that describe the self-similarity and the degree of fractality present in mammogram images. The self-similarity is an inherent property in a number of high-frequency biomedical signals and images. Wavelets, which are local and adaptive functional bases, are suitable for assessing the degree of self-similarity in such data (Vidakovic, 1999). The literature on assessing the scaling exponents is rich and the monograph Doukhan et al. (2003) provides a comprehensive overview. Diagnostics of breast cancer based on scaling measures of mammograms obtained with orthogonal wavelet transform (DWT) and linear regression can be found in Nicolis et al. (2011) and
Roberts et al. (2017). For the same task, multifractal spectral tools have been used in Ram\'irez and Vidakovic (2013),
while the complex wavelets have been utilized by Jeon et al. (2015).

Extending on the aforementioned results, we develop a wavelet spectral scaling estimation method based on the non-decimated wavelet transform (NDWT). The NDWT provides two features that improve on the scaling estimation: First, the NDWT as a redundant transform, yields wavelet coefficients at a maximal sampling rate independently of the multiresolution level. Thus, we obtain the maximum number of wavelet coefficients at all levels, which improves the   stability and accuracy of estimation. Second, the size of an input signal is maintained at each resolution level. This enables us to localize wavelet coefficients corresponding to a region of interest (ROI) at any level in the wavelet domain. We highlight such features in a simulation study and an application for breast cancer detection with mammogram images. In addition, non-decimated transforms do not require dyadic size inputs, which is a constraint for wavelet transforms that decimate.

The remainder of this paper is organized as follows. Section \ref{sec:back} provides background information and a theoretical overview and introduces novel spectral tools in the NDWT-domain. Section \ref{sec:simul} compares the performance of the NDWT-based scaling estimation method to that obtained by the conventional method. Section
\ref{sec:mammo} applies the proposed method to breast cancer detection using digital mammogram images. The concluding section summarizes the results and provides a discussion and final remarks.

\section{Background \label{sec:back}}
To provide a pertinent background of this study, we first overview fractional Brownian motion in one and two dimensions as theoretical models for data/images that scale. This is important for the calibration of the proposed methods since for
the fractional Brownian motion the regularity is a defining index and scaling properties are known exactly.
Next, we review a NDWT, which is a choice transform for mapping an input signal from the acquisition domain to the multiscale domain. Finally, we connect the wavelet spectrum defined on a hierarchy of coefficient spaces in NDWT domain to the signal scaling and its anisotropy properties.

\subsection{Fractional Brownian Motion as Theoretical Model}
Fractional Brownian motions (fBm) and fractional Brownian fields (fBf) are Gaussian self-similar processes with stationary increments. They scale regularly and represent examples of monofractal objects with a singular scaling index, Hurst exponent $H$. For a mathematical representation, let us denote a path of a one-dimensional (1-D) fBm process with Hurst exponent $H$ as $\{B_H(t),\, t\in \mathbb{R} \}$. By definition of self-similarity of fBm, it holds that $B_H(at)$ is equal in distribution to $a^H B_H(t),~ a \geq 0$. The covariance function of ${B_H(t)}$ is
\begin{equation}
\gamma_{B_H}(t,s) = \mathbb{E}\{B_H(t) B_H(s)\}=\frac{ \sigma^2_H}{2}\left( |t|^{2H} + |s|^{2H} - |t-s|^{2H} \right),
~s,t \in \mathbb{R},
\end{equation}
where $\sigma^2_H = \Gamma(1-2H)\frac{cos(\pi H)}{\pi H}$ and $0 \leq H \leq 1$. Because
$\gamma_{B_H} (t,s)$ is not a function of $|t - s|$ only, the fBm is not a stationary process, which implies that we cannot obtain a spectrum of fBm by direct definition. However, we can indirectly deduce a pseudo-spectrum from the fact that increments of fBm are stationary (Reed et al., 1995),
\ba
S_{B_H}(\omega) \propto |\omega|^{-2H-1},
\ea
where $\omega$ indicates frequency. This definition extends to a 2-D fractional Brownian motion, or equivalently, fractional Brownian field (fBf), $B_H(\bm u)$, where $\bm u$ and $\bm v$ are points in 2-D space $[0,1] \times [0,1]$. The covariance function of $B_H(\bm u)$ is
\begin{equation}
\gamma_{B_H}( \bm u, \bm v ) =\mathbb{E} \{B_H( \bm u )B_H( \bm v )\} = \frac{ \sigma^2_H}{2}[ \; ||\bm u||^{2H} + ||\bm v||^{2H} - ||\bm u-\bm v||^{2H} \;],
\label{eq:fbfcov}
\end{equation}
where $||\cdot||$ represents the Euclidean norm, and $\sigma^2_H=\frac{2^{-(1+2H)} \Gamma(1-H)}{\pi H \Gamma }$. As a result, the relationship between the power spectrum and Hurst exponent $H$ is
\ba
S_{B_H}(\omega) \propto ||\omega||^{-2H-2}.
\ea

\subsection{Non-decimated Wavelet Transform}
A non-decimated wavelet transform is a version of wavelet decompositions that offers useful features in a range of applications. We describe algorithmic steps to perform NDWT and discuss its properties. We start with the 1-D NDWT and extend the results into 2-D case.

\begin{figure}[h]
\begin{center}
\scalebox{0.4}{ \includegraphics{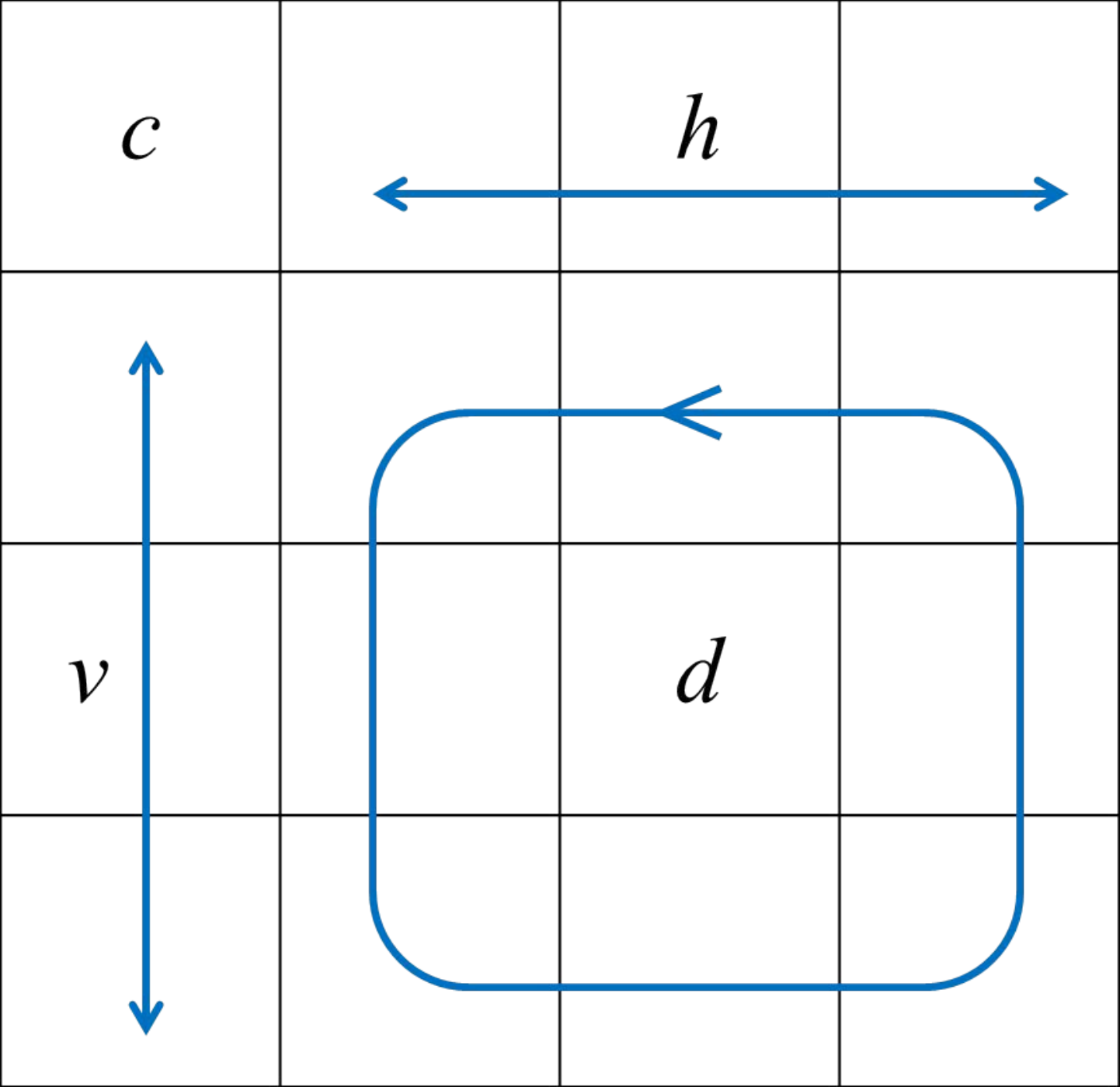}}
\caption{Four types of wavelet coefficients with their locations in the tessellation of a 2-D scale-mixing NDWT
 of depth 3. Coefficients $c$ represent the coarsest approximation, $h$ and $v$ represent the mix of coarse and detail information, and $d$ carry information about details only.}
\label{fig:wavCoef}
\end{center}
\end{figure}

Any square-integrable function, or $L_2(\mathbb{R})$ function, $f(x) $ can be represented as
\ba
f(x)=\sum_{k}c_{j_0, k}\phi_{j_0, k}(x) + \sum_{j\geq j_0}^{\infty} \sum_{  k} d_{j,  k}\psi_{j,k}(x),
\ea
with a scaling and a wavelet functions forming an overcomplete basis with atoms
\ba
\phi_{j_0k}(x)=2^{j_0/2}\phi(2^{j_0}(x - k))\\
\psi_{jk}(x)=2^{j/2}\psi(2^j(x - k)),
\ea
where $x \in \mathbb{R}$, $j$ is a resolution level, $j_0$ is the coarsest resolution level, and $k$ is the location of the atom. Notice that the location shift $k$ is constant at all levels, which indicates that the transform is location invariant. A transformed signal resulting from the NDWT consists of detail and scaling coefficients $d_{jk}$ and $c_{j_0k}$, which can be obtained as inner products
\ba
c_{j_0k} = \langle f(x), \phi_{j_0k}  \rangle\\
d_{jk} = \langle f(x), \psi_{jk}  \rangle.
\ea
Next, we obtain the 2-D NDWT using 1-D transforms performed along the two dimensions. We consider the NDWT of a 2-D input signal $f(x,y) \in L_2(\mathbb{R}^2)$. The tensor product of 1-D wavelet bases is a standard way of generating multidimensional multiresolution analysis. According to this standard approach, for 2-D case, the scaling and wavelet functions are
\ba
\phi_{(j_{01}, k_1)} (x) \phi_{(j_{02}, k_2)} (y) & = & 2^{(j_{01}+j_{02})/2} \phi(2^{j_{01}} (x - k_1)) \phi(2^{j_{02}} (y - k_2)),\\
\phi_{(j_{01}, k_1)}(x) \psi_{(j_{2},k_2)}(y) & = & 2^{(j_{01}+j_2)/2} \phi(2^{ j_{01}} (x - k_1))\psi(2^{ j_2} (y - k_2)),\\
\psi_{(j_1, k_1)} (x) \phi_{( j_{02}, k_2)} (y) & = & 2^{(j_1+j_{02})/2} \psi(2^{ j_1} (x - k_1))\phi(2^{ j_{02}} (y - k_2)),\\
\psi_{(j_1, k_1)} (x)\psi_{(j_2, k_2)} (y) & = & 2^{(j_1+j_2)/2} \psi(2^{ j_1} (x - k_1))\psi(2^{ j_2} (y - k_2)),
\ea
where $(k_1,k_2)$ is the location of atoms, $j_{01}, j_{01}\leq j_1 $ and $j_{02}, j_{02}\leq j_2 $ are the coarsest resolution level for column and row decompositions of $f(x,y)$, respectively. As a result, wavelet coefficients of $f(x,y)$ from the NDWT are
\ba
c_{(j_{01},j_{02}; \bm k)} & = &  \iint   f(x,y)
\phi_{(j_{01}, k_1)} (x) \phi_{(j_{02}, k_2)} (y) \; dxdy, \\
h_{(j_{01},j_{2} ; \bm k)} & = &  \iint   f(x,y)
\phi_{(j_{01}, k_1)}(x) \psi_{(j_{2},k_2)}(y) \; dxdy,\\
v_{(j_{1},j_{02}; \bm k)} & = &  \iint   f(x,y)
\psi_{(j_1, k_1)} (x) \phi_{( j_{02}, k_2)}(y) \; dxdy,\\
d_{(j_1,j_2;\bm k)} & = &  \iint   f(x,y)
\psi_{(j_1, k_1)} (x)\psi_{(j_2, k_2)} (y) \; dxdy,
\ea
where $\bm k=(k_1,k_2)$. Figure \ref{fig:wavCoef} illustrates the location of such coefficients in the wavelet domain.

A constructive method to define an NDWT uses filter operators but here we use series-based representations so that we can connect wavelet coefficients of a signal with its scaling. Interested readers can refer to Nason and Silverman (1995), Vidakovic (1999), and Percival and Walden (2006) for alternative definitions of NDWT.

\subsection{Scaling, Anisotropy, and Wavelet Spectrum}
\begin{figure}[h]
\centering
\scalebox{0.3}{\includegraphics{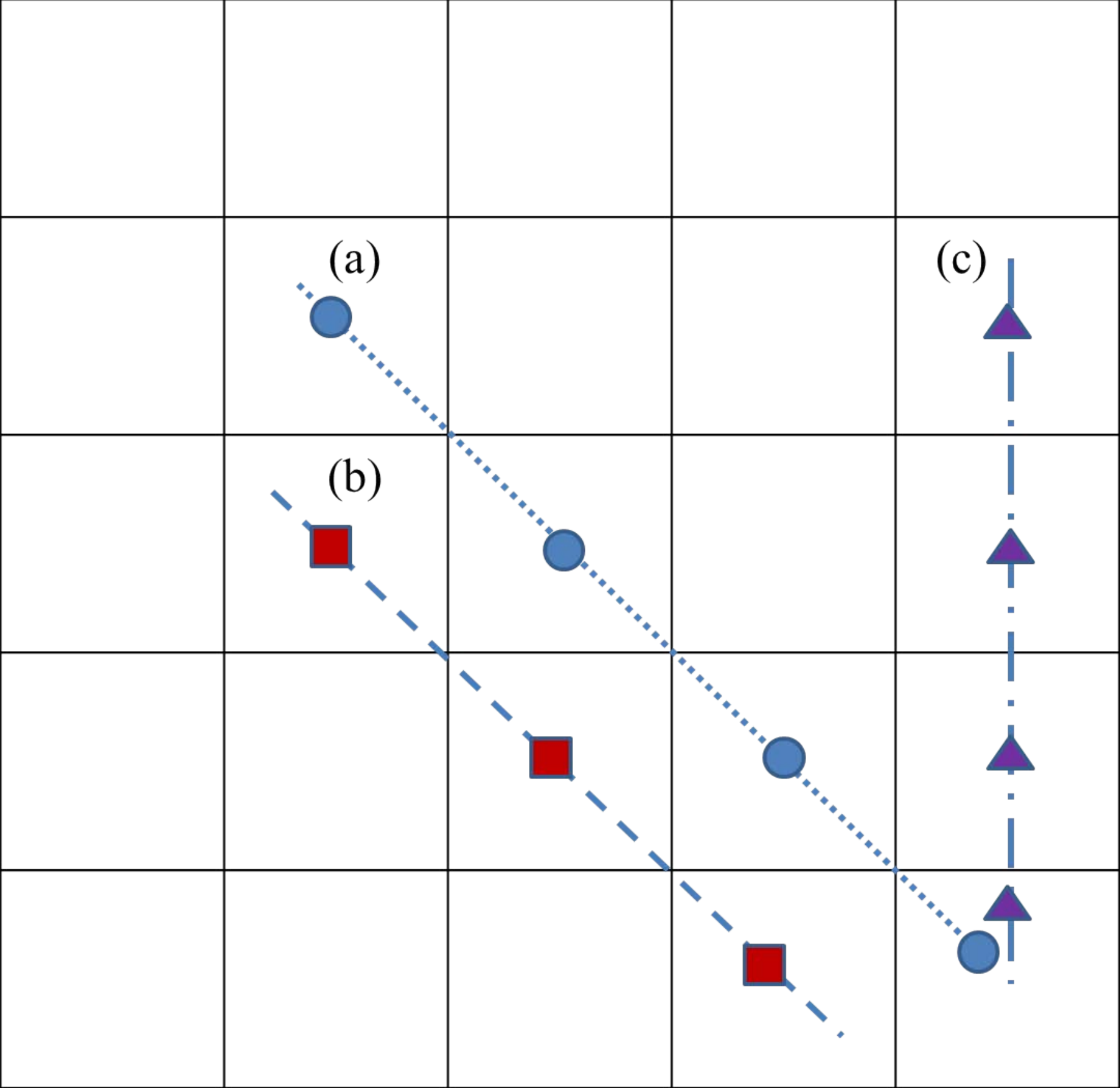}}
\caption{Three hierarchies of detail spaces in the tessellation of 2-D scale mixing NDWT of depth 4,
where
detail spaces are indexed with the pair $(j_1,j_2)$. (a) A main diagonal hierarchy whose scale indices satisfy $j_1=j_2$, (b) a hierarchy in which the scale  index $j_2$ is fixed to the finest resolution level, and (c) a hierarchy where the scale indices satisfy $j_1= j_2+1.$}
\label{fig:hierarchy}
\end{figure}

To quantify characteristics within 2-D inputs with a wavelet spectrum, we obtain two types of descriptors: scaling and asymmetry measures. Defined in a time/scale domain, a wavelet spectrum represents the distribution of energies within an original signal along the range of scales (i.e., resolution levels). In the wavelet jargon, the term ``energy'' stands for a squared wavelet coefficient. For each 2-D resolution level indexed by pair $\bm j=(j_1,j_2)$, $\overline{|d_{\bm j}|^2}$ represents the average level energy.  The wavelet spectra refers to a sequence of logarithms of average level energies along a hierarchy that can be selected in various ways. Figure \ref{fig:hierarchy} demonstrates three possible hierarchies in a tessellation of 2-D scale-mixing NDWT. In this study, we focus on the main diagonal hierarchy whose 2-D scale indices coincide, i.e., $j_1=j_2$. We denote a set of levels, which belong to the main diagonal hierarchy as $\bm j_s=(j,j), \; \text{where } \max(j_{02},j_{01}) \leq j \leq J-1$.
Wavelet coefficients obtained from NDWT possess spatial location invariance across the level spaces. Thus, once the ROI in the original signal is selected, one can easily identify wavelet coefficients
in each resolution level
that correspond to the ROI. Therefore, when calculating a wavelet spectrum, one can use either all wavelet coefficients or only the portion of coefficients corresponding to a ROI. Such local spectra are natural for NDWT, unlike the orthogonal transforms that decimate.
Right panel in Figure \ref{fig:scale} exemplifies this feature.

\begin{figure}[tb]
\begin{center}
\scalebox{0.4}{ \includegraphics{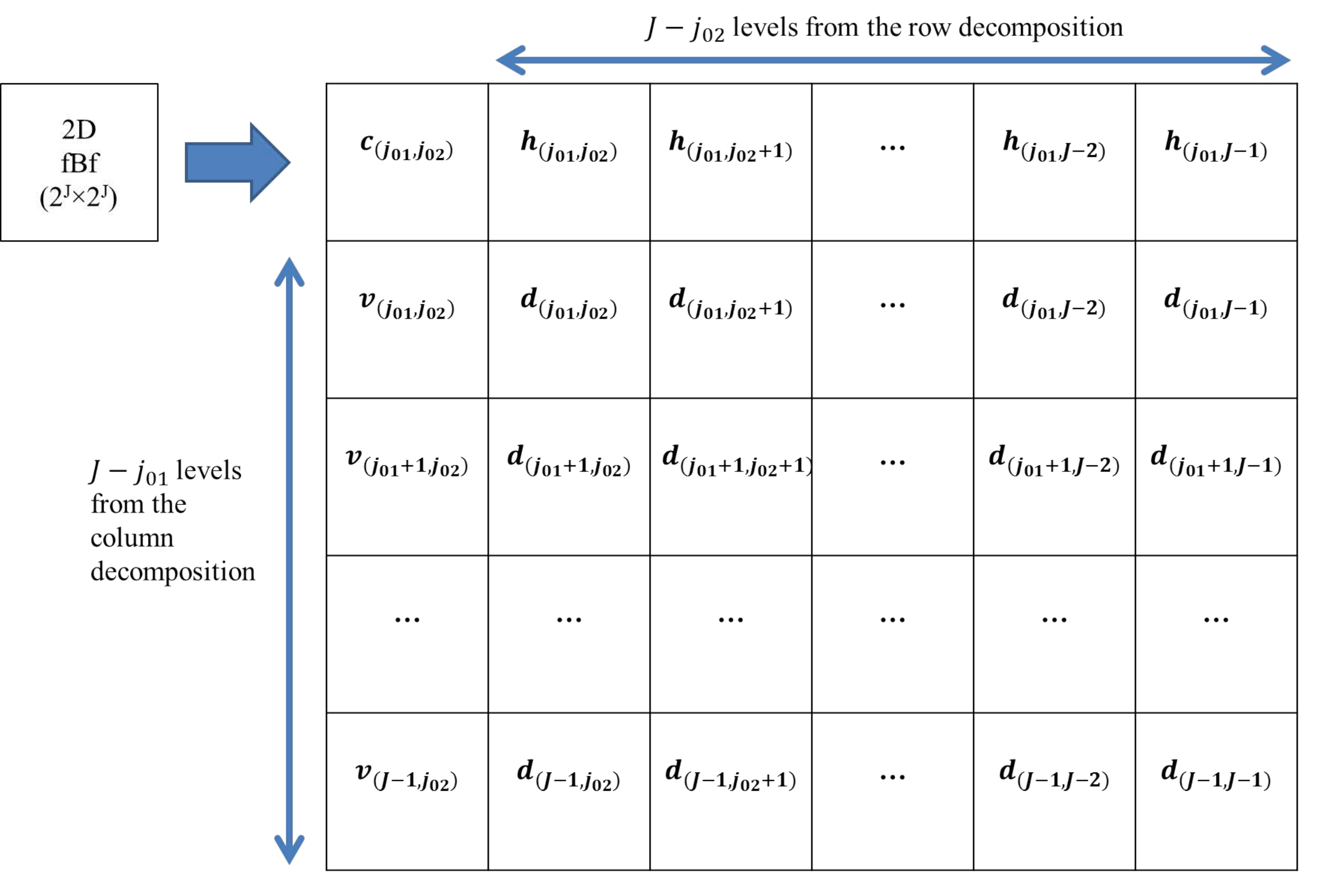}}
\caption{The location of wavelet coefficients that belong to various resolution levels when we perform 2-D scale-mixing NDWT to an image of the size $(2^J \times 2^J)$ with the lowest resolution levels for columns
and rows of $j_{01}$ and $j_{02}$, respectively. }
\label{fig:coefficients}
\end{center}
\end{figure}

\begin{figure}[tb]
\begin{center}
\scalebox{0.4}{ \includegraphics{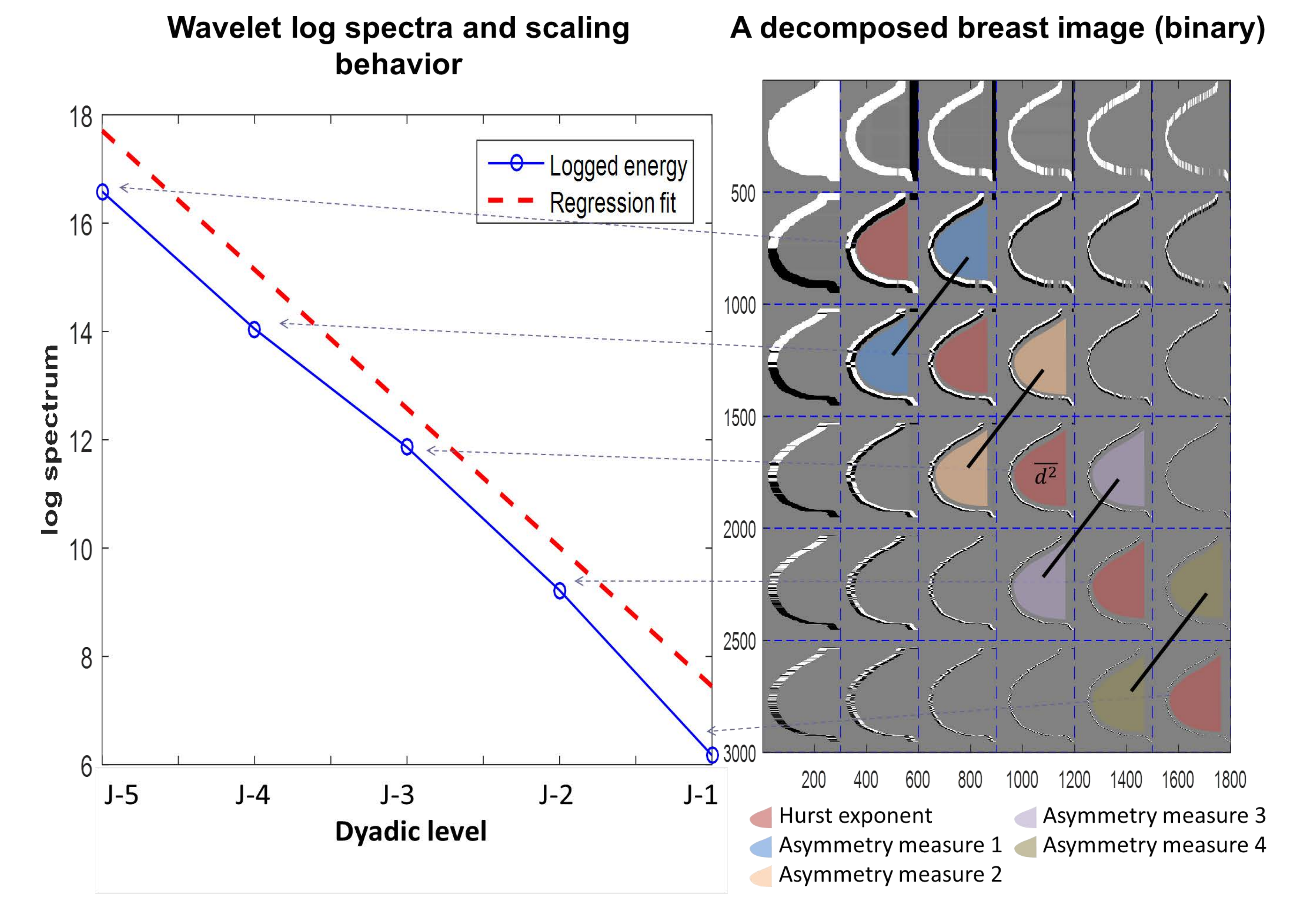}}
\caption{Left panel: Wavelet spectrum of log average energies across the scales represented by a solid line and its linear fit represented by a dotted line. Right panel: A 2-D  NDWT of depth 5 of a ``mask'' image with main diagonal hierarchy and pairs of subspaces for anisotropy calculations. }
\label{fig:scale}
\end{center}
\end{figure}

For the estimation of the first descriptor, scaling, we measure the rate of average energy decrease along $\bm j_s$ from the main diagonal hierarchy. When this decrease of energy is regular, it can be connected to a degree of self-similarity in signals and defines a commonly used scaling index, Hurst exponent $H$. The relationship between energies in the wavelet spectra and Hurst exponent $H$ is captured by a simple equation. To explain the equation in detail, we consider one example with a fBf ${B_H( x,y)}$ of the size $(2^{J} \times 2^J)$. We perform 2-D scale-mixing NDWT to ${B_H( x,y)}$ with the lowest resolution levels for columns and rows as $j_{01}$ and $j_{02}$, respectively. Figure \ref{fig:coefficients} demonstrates the resulting resolution space and wavelet coefficients yielded from this transform.
The wavelet spectrum from the main diagonal hierarchy  is defined as a set of pairs
\be
\label{eq:pairs}
 \left(j,~ \log_2 \left(\overline{|d_{ (j,j) ; \bm k}|^2}\right)~\right),
\ee
where $(j,j)\in \bm j_s$ and $\max (j_{02},j_{01}) \leq j < J$. Wavelet coefficients $d_{ (j,j) ;\bm k}$ are approximately independent and identically distributed Gaussian random variables with  zero mean and a variance dependent on level $j$  (Heneghan et al., 1996). The expected energy at each level in the main diagonal hierarchy is
\begin{align}
E\big[| d_{ (j,j) ; \bm k }|^2\big] & = 2^{2{j }} \iint   \psi \big( 2^{j }(\text{\textbf v} - \bm k) \big) \,  \psi \big( 2^{j }(\text{\textbf u} - \bm k) \big) E\big[B_H(\text{\textbf v})B_H(\text{\textbf u})\big] \,d\text{\textbf v} d\text{\textbf u}\nonumber\\
& = \frac{\sigma^2_H}{2}V_{\psi } 2^{-(2H+2)j },
\label{eq:energy}
\end{align}
where $(j,j)\in \bm j_s$, \textbf u,\textbf v $\in\mathbb{R}^2$, $\bm k =(k_1,k_2)$, and
\begin{align*}
V_{\psi } = -  \iint \psi(\bm p +\bm q) \psi (\bm q)|\bm p|^{2H} d\bm p d\bm q.
\end{align*}
Expression $V_{\psi}$ is independent of scale $j$, but dependent on wavelet function $\psi$ and Hurst exponent $H$. Details on derivation of (\ref{eq:energy}) are deferred to Appendix. Taking the logarithm on both sides in (\ref{eq:energy}) yields
\be
\log_2 E\big[|d_{ (j,j) ;\bm k}|^2] =-(2H + 2)j  + C,
\label{eq:logE}
\ee
where $(j,j)\in \bm j_s$ and $C$ is a constant that does not depend on $j$ but depends on  wavelet function $\psi$ and $H$. Figure \ref{fig:scale} provides a graphical explanation for how a wavelet spectrum is formed for the estimation of scaling.
In the left panel of Figure \ref{fig:scale}, a wavelet spectrum of log average energies across the scales represented by a solid line and its linear fit, represented by a dotted line, are shown.
 The right panel of Figure \ref{fig:scale} shows a 2-D  NDWT of depth 5. Red marked areas represent wavelet coefficients used for scaling estimation from the main diagonal hierarchy. The left top image in the matrix represents the coarsest features of an original image.
Note that for calculation of the scaling descriptors, we use only wavelet coefficients located in the colored areas, i.e., corresponding
to the area with tissue, since we are interested in local spatial characteristics. At each resolution level, we can readily identify wavelet coefficients that correspond to the selected ROI in the original image.
This spatial location invariance across the scales is distinctive feature of NDWT's and enables the construction of local spectra.

Once the slope $\hat \beta$ in the linear fit of pairs in (\ref{eq:pairs}),
according to (\ref{eq:logE}), is found, the Hurst exponent is estimated as $\hat H = - \hat \beta/2 - 1.$

For the estimation of the second descriptor $A_j$, a degree of horizontal/vertical anisotropy, we calculate the asymmetry ratio of two average energies that are adjacent to the main diagonal hierarchy, for which scale indices differ
by 1. Thus, an asymmetry measure at level $j$ is
$A_j = \overline{|d_{(j+1,j)}|^2}/\overline{|d_{(j,j+1)}|^2}$, where $(j,j)\in \bm j_s$. If an input image exhibits isotropy in horizontal and vertical directions, the ratio is close to 1. Deviations from this ratio could be informative.
Figure \ref{fig:scale} visually describes the linked pairs of subspaces used for calculation of four asymmetry measures.

\section{Simulated Cases \label{sec:simul}}
In a simulation study, we compare the performance of NDWT- and DWT-based scaling estimation methods. We simulate three sets of 500 2-D fBm signals of size $2^9 \times 2^9$ with Hurst exponents 0.3, 0.5, and 0.7, respectively. Next, we transform each signal into $8$ multiresolution subspaces for both row and column decompositions with a 2-D scale-mixing NDWT based on four wavelets:  Daubechies 6, Symmlet 8, Coiflet 6, and Haar. The wavelets are indexed by number of filter components and not by the number of vanishing moments. We found that the estimation performance is robust with respect to choice of wavelet filter.
In estimating the scaling, we take logarithm on part of the main diagonal wavelet spectrum which includes level $(j,j)$ where $j=2,3, \dots , 6$, and then fit a linear regression model on log average level energies.
The slope of this linear regression leads the estimator of Hurst exponent. We evaluate the performance of both NDWT- and DWT- based estimation methods, by comparing their means, variances, and biases.

Unlike the decimated case in which the number of coefficients differs at each level, and correct
linear fitting procedure involves weighted regression Veitch and Abry (1999), here the ordinary least square
(OLS) regression provides theoretically correct and satisfactory fit.
Because of redundancy, dependence among neighboring wavelet coefficients within the same
level is much more pronounced than in the case of orthogonal wavelets.
Although this dependence is not biasing estimators, the variances of estimators are affected.
Another factor that influences variance is the choice of wavelet basis, and more local wavelet bases are preferred.

\begin{figure}[!htb]
\begin{center}
\includegraphics[scale=0.7, clip]{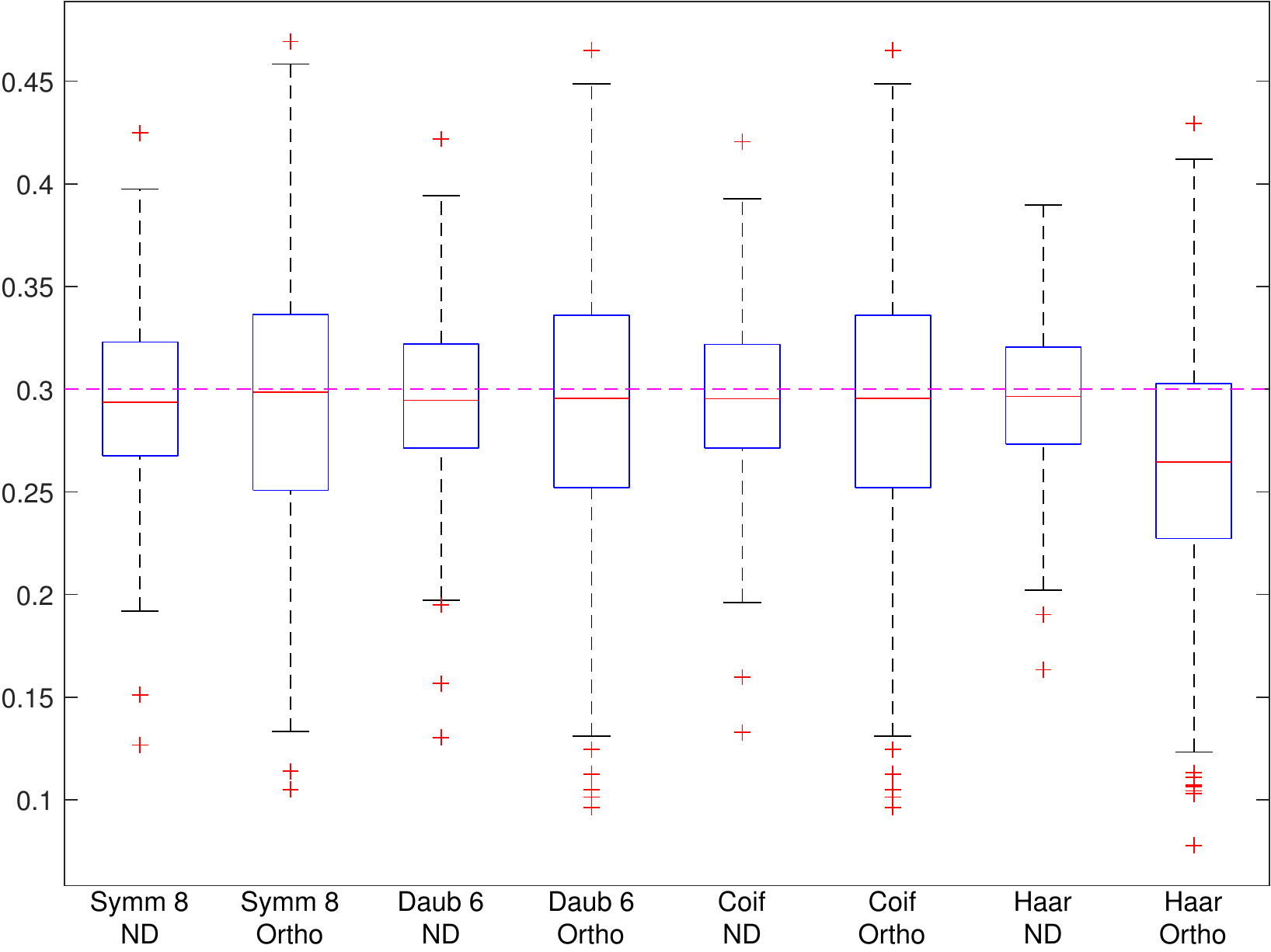}
\caption{Boxplots of $\hat H$ from 300 simulations of 2-D fBm's when $H=0.3$ with various wavelet bases
 and non-decimated and orthogonal transform.}
\label{fig:H3}
\end{center}
\end{figure}

\begin{figure}[!htb]
\begin{center}
\scalebox{0.7}{ \includegraphics{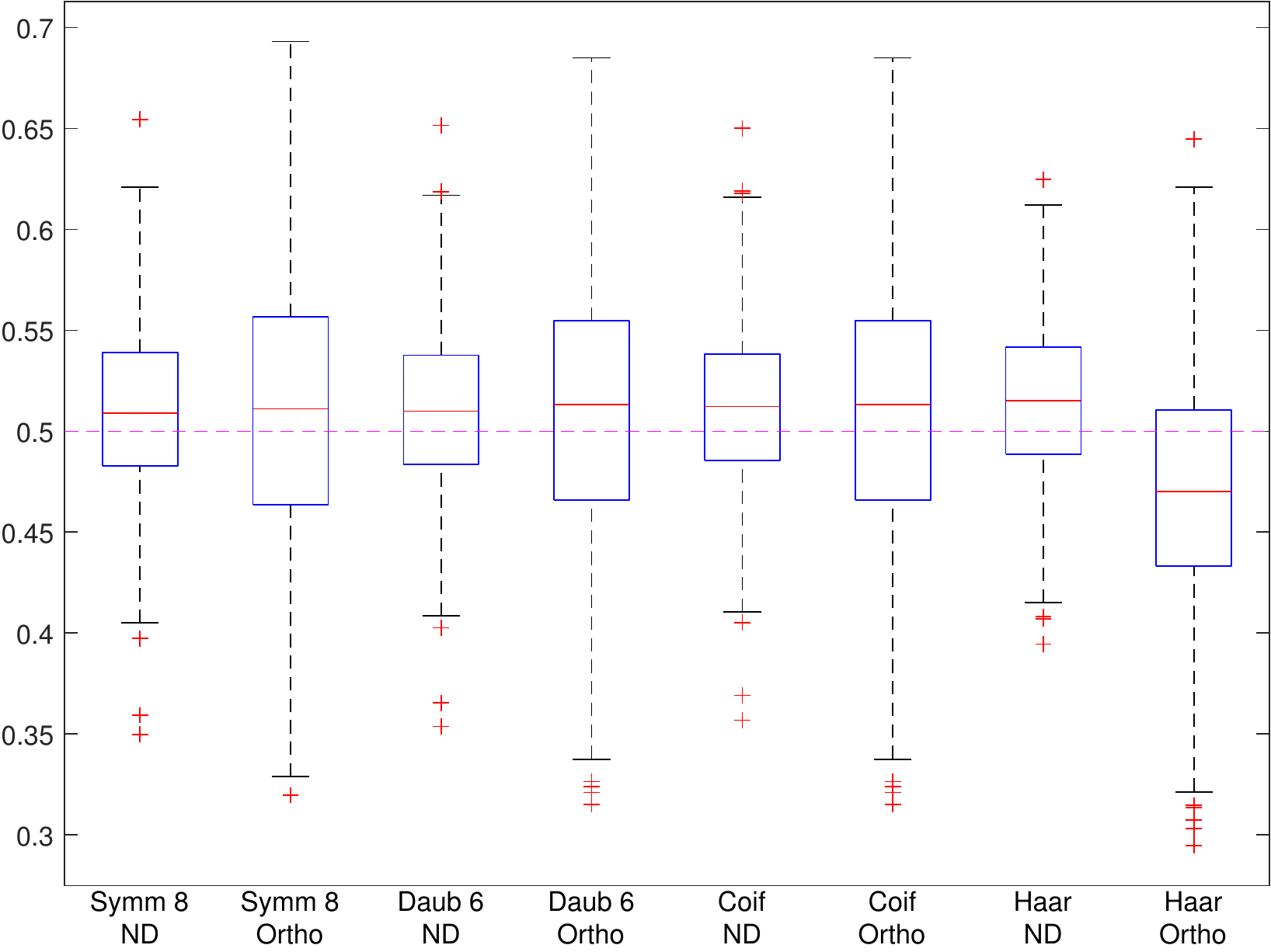}}
\caption{As in Figure \ref{fig:H3}, but for $H=0.5.$}
\label{fig:H5}
\end{center}
\end{figure}

\begin{figure}[htb]
\begin{center}
\includegraphics[scale=0.7, clip]{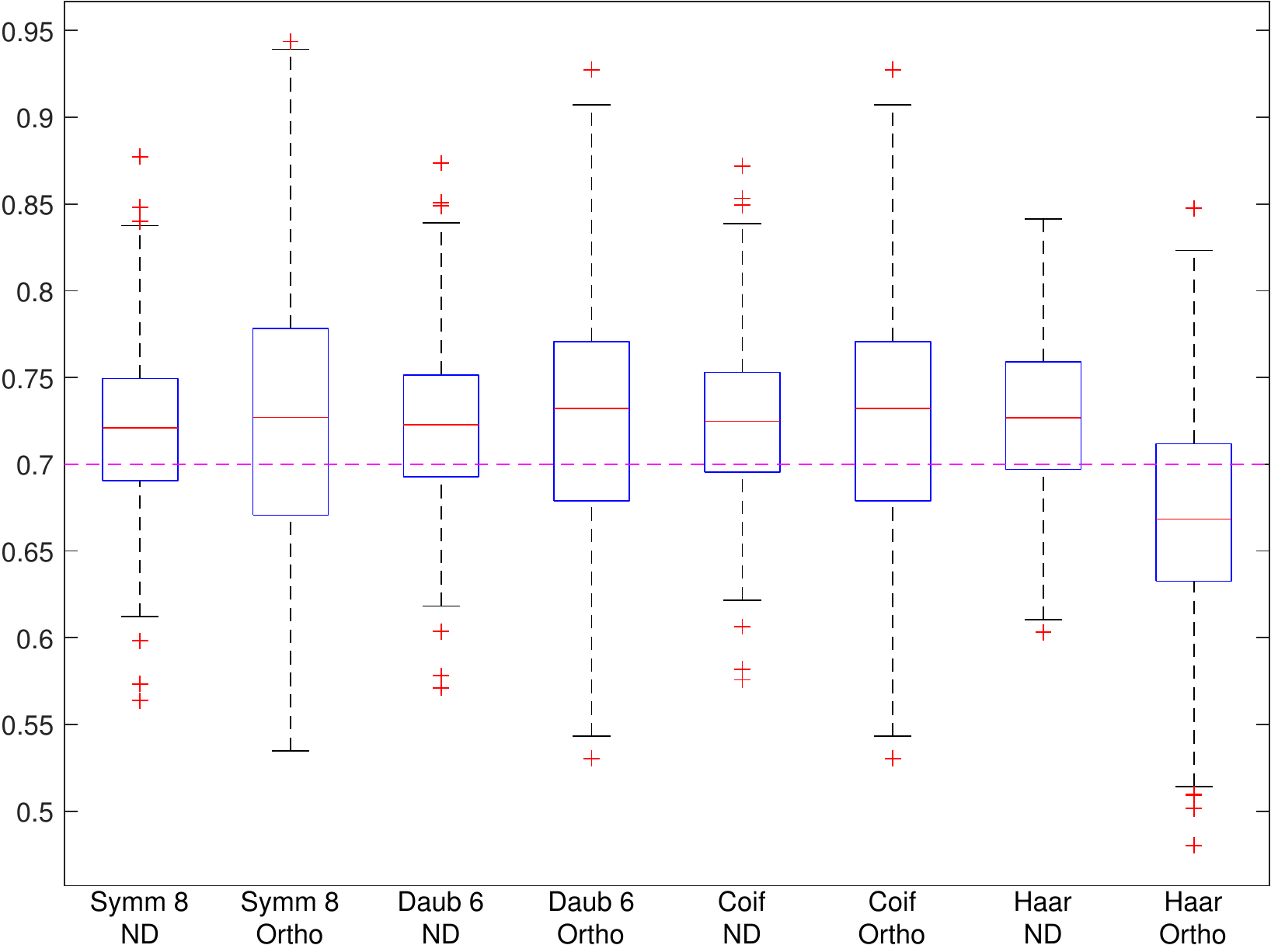}
\caption{As in Figure \ref{fig:H3}, but for $H=0.7.$}
\label{fig:H7}
\end{center}
\end{figure}

Tables \ref{tab:03}-\ref{tab:07}
  summarize the estimation results under various settings. An average of $\hat H$, its mean-square error, variance, and bias-squared,
based on 300 2-D fBm's when true Hurst exponents, $H=0.3, 0.5$ and 0.7, are provided.
 Symmlet 8, Daubechies 6, Coiflet 6, and Haar wavelet bases are used in
non-decimated and orthogonal versions of wavelet spectra.
Figure \ref{fig:H3}-\ref{fig:H7} show box-and-whisker plots of simulations described in
Tables \ref{tab:03}-\ref{tab:07}.

\begin{table}
\centering
\begin{tabular}{ c| c c| c c |c c| c c }
  Wavelets & \multicolumn{2}{c}{Symmlet 8} \vline & \multicolumn{2}{c}{Daubechies 6}\vline & \multicolumn{2}{c}{Coiflet 6}\vline & \multicolumn{2}{c}{Haar} \\
  \hline
   &  ND & Ortho &  ND & Ortho & ND & Ortho & ND & Ortho \\  \hline
Mean &	0.2946 &	0.2945 &	0.2955 &	0.2939 &	0.2963 &	0.2939 &	0.2959 &	0.2632 \\
MSE &	0.0017 &	0.0043 &	0.0016 &	0.0041 &	0.0015 &	0.0041 &	0.0012 &	0.0047 \\
Variance &	0.0017 &	0.0043 &	0.0015 &	0.0041 &	0.0015 &	0.0041 &	0.0012 &	0.0034 \\
Bias & 2.2E-5	& 1.3E-5& 	1.4E-5& 	2.2E-5& 	7.9E-6& 	2.2E-05& 	1.2E-5 & 0.0013 \\
\end{tabular}
\caption{An average of $\hat H$, its mean-square error, variance, and bias-squared,
based on 300 2-D fBm's when true $H=0.3$ obtained by various wavelet-bases and transform choices, i.e., non-decimated and orthogonal.  \label{tab:03} }
\end{table}

\begin{table}
\centering
\begin{tabular}{ c| c c |c c| c c |c c }
  Wavelets & \multicolumn{2}{c}{Symmlet 8}\vline & \multicolumn{2}{c}{Daubechies 6} \vline& \multicolumn{2}{c}{Coiflet 6}\vline & \multicolumn{2}{c}{Haar} \\
  \hline
    &  ND & Ortho &  ND & Ortho & ND & Ortho & ND & Ortho \\  \hline
Mean &	0.5115 &	0.5096 &	0.5109 &	0.5112 &	0.5109 &	0.5125 &	0.4703 &	0.5153 \\
MSE &	0.005 &	0.002 &	0.0046 &	0.0019 &	0.0046 &	0.0018 &	0.0044 &	0.0018 \\
Variance &	0.0049 &	0.0019 &	0.0045 &	0.0018 &	0.0045 &	0.0017 &	0.0035 &	0.0016 \\
Bias &	0.0001 &	8.5E-5&	9.9E-5&	0.0001&	9.9E-5 &	0.0001 &	0.0009 &	0.0002 \\

   \\
\end{tabular}
\caption{As in Table \ref{tab:03} but for $H=0.5.$  \label{tab:05}}
\end{table}

\begin{table}
\centering
\begin{tabular}{ c| c c |c c| c c |c c }
  Wavelets & \multicolumn{2}{c}{Symmlet 8} \vline& \multicolumn{2}{c}{Daubechies 6} \vline & \multicolumn{2}{c}{Coiflet 6}\vline & \multicolumn{2}{c}{Haar} \\
  \hline
   &  ND & Ortho &  ND & Ortho & ND & Ortho & ND & Ortho \\    \hline
Mean &	0.7212 &	0.727 &	0.7279 &	0.6688 &	0.7237 &	0.7267 &	0.7256 &	0.7267 \\
MSE &	0.0026 &	0.0065 &	0.0028 &	0.0045 &	0.0026 &	0.0058 &	0.0026 &	0.0058 \\
Variance &	0.0022 &	0.0058 &	0.002 &	0.0035 &	0.002 &	0.0051 &	0.002 &	0.0051 \\
Bias &	0.0004 &	0.0007 &	0.0008 &	0.001 &	0.0006 &	0.0007 &	0.0006 &	0.0007 \\
\end{tabular}
\caption{As in Table \ref{tab:03} but for $H=0.7.$  \label{tab:07} }
\end{table}

Because NDWT produces the maximum number of wavelet coefficients at each resolution level, we are able to obtain more accurate estimates of energies with more coefficients to average at each level. Thus, the NDWT-based method yields estimators with lower mean squared errors compared to the DWT-based method.
As it can be seen in Tables \ref{tab:03}-\ref{tab:07} and Figures \ref{fig:H3}-\ref{fig:H7}, the empirical variances are influenced by the choice of wavelet.
The redundancy of NDWT turned out to be beneficial  despite some negative effect of increased levelwise dependence among the coefficients.

\section{Application in Mammogram Diagnostic \label{sec:mammo}}
We apply the 2-D scale-mixing NDWT-based method to digital mammograms with the goal of identifying wavelet features suggestive of breast cancer.
\subsection{Source of Data}
\begin{figure}[!htb]
\begin{center}
\scalebox{0.75}{ \includegraphics{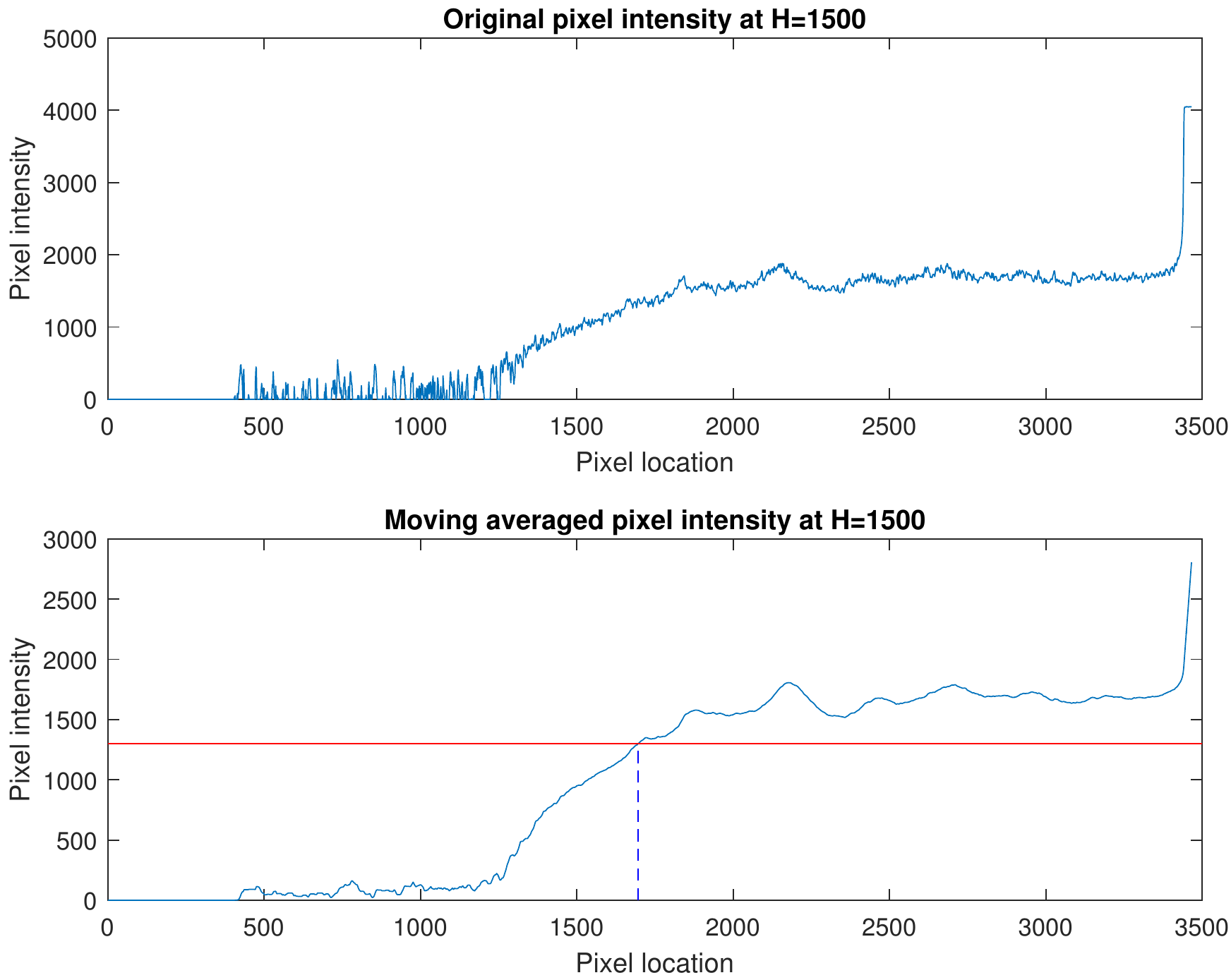}}
\caption{The top panel depicts an exemplary row of pixel intensities
from an original image
that exhibits fluctuations. The bottom panel shows the pixel intensity smoothed by a moving average, from which we are able to identify a single boundary pixel.}
\label{fig:edge}
\end{center}
\end{figure}

We obtain the mammographic images from Digital Database for Screening Mammography (DDSM) at the University of South Florida  Heath et al. (2001). All cases examined had biopsy results which served as ground truth. University researchers used a HOWTEK scanner at the full 43.5-micron per pixel spatial resolution to scan 45 mammograms from patients without breast cancer and 79 from patients with confirmed breast cancer.  Each case contains two images of each breast in craniocaudal (CC) and mediolateral oblique (MLO) projections. From these images, we obtain 124 CC projection images and identify background tissue area of a breast in each mammographic image so that we can extract wavelet features that indicate the health conditions of patients based on the identified tissue areas without affects of background non-breast area. Figure \ref{fig:edge} shows the mammogram and the mask image, which is a binary image that takes value 1 if the location belongs to a breast tissue area or 0 otherwise. In a subsequent classification process, we use the mask image to filter out numerical values (i.e., wavelet coefficients) from NDWT that are irrelevant to defining self-similar properties of breast tissues.

\subsection{Diagnostic Classification}
For breast cancer diagnostics, we performed 2-D scale-mixing NDWT of depth 6 for each mammogram. The mammograms had various sizes and the location of a breast tissue area within a mammogram also varied. As we were interested exclusively in the scaling characteristics of the breast tissue, we first identified the wavelet coefficients, which belong to the breast tissue area using a two-step process. We began by orienting all mammograms from left to right, so that a breast tissue area starts at the right-hand side of mammogram. Then, for each row, we defined the boundary of the breast tissue area. This was done by an algorithm that selected the left-most-pixel for which the average intensity of the 64 subsequent pixels decreased below a predefined threshold $\lambda.$
 We averaged a sequence of pixel intensities so that noisy fluctuations among pixel intensities
in a row are smoothed, to prevent multiple boundary points in a single row of pixels.
Once the boundary point in each row of pixels was identified, we
adjust for the ``cone-of-influence'' effect typical for all wavelet transforms.
The  cone-of-influence effect refers to a blurring effect of  wavelet filters when
applied in a sequential manner, like in Mallat's algorithm. Because of this blurring effect,
a local feature of a signal propagates along the multiresolution spaces in a shape of a cone.
The longer the wavelet filter, the wider the cone.
To eliminate this effect, that is, to eliminate influence of non-tissue pixels on the local wavelet
spectra, we
shift $m$ pixel-locations to the right from the original boundary to form an updated boundary.

To emphasize locality, we used Haar wavelet, which produces the most narrow cone.
The maximum length of a Haar wavelet filter convolved over
 the $6$-level non-decimated decomposition is bounded by $2^6$ which is an approximation to the maximal width of the cone.
 For comparisons and robustness assessment, in further analysis we selected three $m$ values, $0$, $2^6$, and $2^7$.

  Based on the boundary construction rules for each row, we form a 0-1 image (mask) so that we can select wavelet coefficients corresponding to the ROI through a simple multiplication.

  An entry at row $i$ and column $j$ of a binary image (mask) $\bm \mu$ is
\ba
\mu_{ij} = \left\{ \begin{array}{ll}
      1,~~~  \frac{1}{m} \sum_{k = j + 1}^{j+m} I_{ik} > \lambda \mbox{ ~~and~~} j \geq m+1 \\
      0,~~~  \mbox{else}
      \end{array}
      \right.
      \ea
where $I_{ik}$ indicates a pixel intensity in a mammogram at position $(i,k)$, and $m$ the length of a shift.  The mask $\bm{\mu}$ has the same size as the mammogram.
 The entry of $\bm{\mu}$ at location $(i,j)$ is an indicator that is 1 if pixel $(i,j)$ belongs to a breast tissue region and is unaffected by the cone of influence, and 0 otherwise. We perform Hadamard (element-by-element) multiplication of a mask
image and wavelet coefficients at each resolution level. With such multiplication, only wavelet coefficients that belong to the breast tissue region at each level are selected. From those selected wavelet coefficients, we find five descriptors: a scaling measure and four asymmetry measures, as in Panel (b) of Figure \ref{fig:scale}. \\
As we discussed, the asymmetry measure, compares horizontal vs vertical isotropy of in breast tissue.

The scaling descriptor, is calculated by the equation (\ref{eq:logE}) from coefficients in the main diagonal hierarchy, $\overline{|d_{(J-5,J-5)}|^2}$, $\overline{|d_{(J-4,J-4)}|^2}$, $\overline{|d_{(J-3,J-3)}|^2}$, $\overline{|d_{(J-2,J-2)}|^2}$, and $\overline{|d_{(J-1,J-1)}|^2}$. The choice of diagonal hierarchy provides the most information about the regularity of breast tissue (Nicolis et al, 2011).

The asymmetry measures are the energy ratios  of two adjacent levels to
 the main diagonal hierarchy $(j,j)$, indexed by $(j,j-1)$ and $(j-1,j)$ for $ j= J-4, J-3, J-2$, and $J-1$.
 For example, at scale $j=J-3$, an asymmetry measure is defined as $\overline{|d_{(J-3,J-4)}|^2}$/$\overline{|d_{(J-4,J-3)}|^2}$.

We found that the five scales
 of finest detail  were most effective in classification of the health conditions of patients because
 disease signatures are mostly captured in subtle variations within the tissue area. Note that the energy at each level is calculated with only the wavelet coefficients located in the breast tissue area.
\begin{figure}[!htb]
\begin{center}
\scalebox{0.8}{ \includegraphics{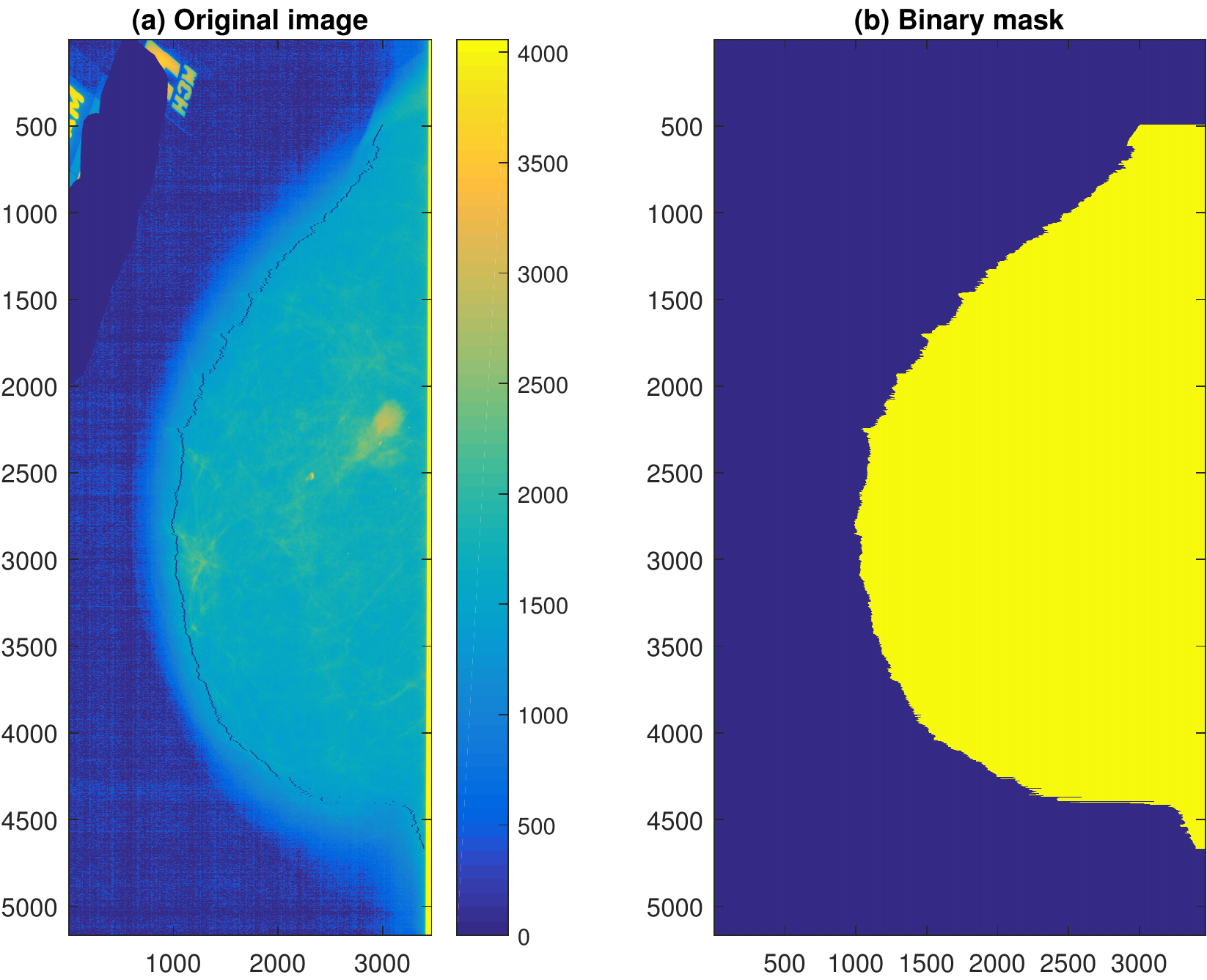} }
\caption{An original mammogram (a) and a binary mask (b) indicating the domain of wavelet coefficients used in the analysis. The black line in the original image represents the boundary detected by the algorithm.
Panel (b) shows the mask image in which white corresponds to 1 and black to 0.}
\label{fig:mammo}
\end{center}
\end{figure}

We use the obtained five features of all mammograms as inputs to three classifiers: logistic regression, support vector machine, and random forest algorithm. In each iteration, we use four-fold cross validation, which randomly divides the data into four sets and then uses three sets as training data and the remaining set as test data. We repeat this random division of training and testing data sets 200 times, and the report  prediction accuracies averaged over these 200 repetitions.

Because we have unequal numbers of cases for cancer and control, which are 45 cases and 79 cases, respective, we perform a random undersampling in selecting cancer cases  (He et al., 2009). Thus, in each iteration, we have the same number of cases (i.e., 45) for both cancer and control in building one classifier and we use four-fold cross validation, which randomly divides the data into four sets and then uses three sets as training data and one remaining set as test data. We repeat this random division of training and testing data sets $1,000$ times, and thus the reported prediction accuracies are averaged over these $1,000$ repetitions.

We present and compare the performance of classification in terms of sensitivity, specificity, and the overall classification accuracy, which are shown in Tables
\ref{tab:sensitivity}, \ref{tab:specificity}, and   \ref{table:accuracy}.
\begin{center}
\begin{tabular}{ c c c c c c}
$m$ value   & 	 \vline & 	Logistic regression   & 	SVM   & Random forest     \\   \hline
0   & 	 \vline & 	0.7354   & 	0.6811   & 	0.8511     \\
$2^6$   & 	 \vline & 	0.7692 & 	0.7104   & 	0.8721     \\
$2^7$   & 	 \vline & 	0.7703 & 	0.7213   & 	0.8739     \\
\end{tabular}
\captionof{table}{Sensitivity with three classifiers. All algorithms show strong diagnostic power in identifying cancerous mammograms. }
\label{tab:sensitivity}
\end{center}

\begin{center}
\begin{tabular}{ c c c c c c}
$m$ value   & 	 \vline & 	Logistic regression   & SVM & Random forest     \\  \hline
0   & 	 \vline & 	0.6293   & 	0.585   & 	0.585     \\
$2^6$   & 	 \vline & 	0.6642   & 	0.5865   & 	0.5865     \\
$2^7$   & 	 \vline & 	0.6572   & 	0.5954   & 	0.5954     \\
\end{tabular}
\captionof{table}{Specificity with three classifiers.}
\label{tab:specificity}
\end{center}

\begin{center}
\begin{tabular}{ c c c c c c}
$m$ value & \vline & 	Logistic regression  & 	SVM   & Random forest   \\ \hline
0   & \vline & 	0.692   & 	0.6474   & 	0.7975     \\
$2^6$ & \vline & 	0.7264   & 	0.6655   & 	0.8272     \\
$2^7$ & \vline  & 	0.7256   & 	0.6753   & 	0.8335     \\
\end{tabular}
\captionof{table}{Classification accuracy with three classifiers. Random forest algorithm shows the best diagnostic accuracy exceeding 80\%.}
\label{table:accuracy}
\end{center}

\section{Conclusions}
Most existing computer aided breast cancer detection methods focus on identifying markers of breast cancer in specific regions. The diagnostic use of information contained in the background tissue is not evaluated.
This paper relates the degree of self-similarity and anisotropy of patterns in breast tissue areas
of a mammogram to the presence of breast cancer. We develop a 2-D scale-mixing NDWT based method that estimates the degree of scaling behavior and anisotropy of breast background tissue. We first assess the scaling estimation performance of the proposed method in simulated cases with 2-D fBm's. In the simulations, the proposed method yields, on average, scaling estimators closer to the target values and with lower mean square errors. Then, we apply the NDWT method to publicly available mammographic images from University of South Florida (Heath et al., 2001) for the detection of breast cancer.
 The selected classifiers use five descriptors:  one self-similarity measure and four asymmetry measures. Computation of those descriptors benefited from two distinctive characteristics of non-decimated wavelet transforms. First, the redundancy of transform produced estimators with smaller variance without inducing additional bias, and the second, the spatial invariance of the transform enabled calculation of local spectra so that coefficients not corresponding to breast tissue were excluded from the analysis. With the five descriptors described in this paper, we achieved an average diagnostic accuracy in excess of 80\%.

One of the valid criticisms for the clinical use of this methodology is
that the accuracy rate is not high enough. Indeed, this would be the case if the proposed method is to be used
by itself. However, even the classifiers, ``slightly better than flipping a coin," can
improve accuracy when used in conjunction with other independent testing modalities.
In this respect, our findings provide an opportunity for significant improvement
of existing mammogram classification procedures and can assist the radiologist.

\vspace*{0.3in}
\noindent {\bf Acknowledgement.~}
Supported by the National Center for Advancing Translational Sciences of the National Institutes of Health under Award Number UL1TR000454. The content is solely the responsibility of the authors and does not necessarily represent the official views of the National Institutes of Health. The work of Brani Vidakovic was supported in part by Giglio Family Award for Breast Cancer Research.

\begin{center}
{\large  REFERENCES}
\end{center}

\leftskip 0.3truein
\parindent -0.3truein

American Cancer Society. Cancer Facts \& Figures 2021.   American
Cancer Society, Atlanta, GA  


 Doukhan, P.,  Oppenheim, G.,  and   Taqqu, M.S. (2003). {\it Theory and Applications of
Long-range Dependence.} , Birkha\"user, Basel.

 He, H. and Garcia, E.A. (2009). Learning from imbalanced data.
 {\it IEEE Transactions on Knowledge and Data Engineering},
  {\bf 21}, 9, 1263--1284.

 Heath, M.,   Bowyer, K.,    Kopans, D.,   Moore, R., and   Kegelmeyer, W.P. (2001).
The digital database for screening mammography,  In {\it Proceedings of the Fifth International Workshop on Digital Mammography}, M.J. Yaffe, ed., 212--218, Medical Physics Publishing,   ISBN 1-930524-00-5.

 Heneghan, C.,    Lowen, S.B., and    Teich, M.C. (1996). Two-dimensional fractional
Brownian motion: Wavelet analysis and synthesis. In {\it Proceedings of the IEEE Southwest Symposium on Image Analysis and Interpretation,
April 8-9, 1996, San Antonio, TX},
213--217.

  Jeon, S.,  Nicolis, O., and   Vidakovic, B. (2015). Mammogram diagnostics via 2-D
 complex wavelet-based self-similarity measures. {\it S\~ao Paulo
Journal of Mathematical Sciences}, {\bf 8}, 2, 256--284.

 Martin, J.,   Moskowitz, M., and   Milbrath, J. (1979).
  Breast cancer missed by mammography. {\it American Journal of Roentgenology}, {\bf 37}, 2, 142--162.

 Nason. G.P. and Silverman, B.W. (1995). The stationary wavelet transform
and some statistical applications. In {\it Wavelets and Statistics},
Springer, NY, pp 281--299.

Nicolis, O., Ram\'irez-Cobo, P., and   Vidakovic, B.  (2011). 2D wavelet-based
spectra with applications. {\it Computational Statistics \& Data Analysis}, {\bf 55},
738--751.

  Percival, D.B. and  Walden, A.T. (2006). {\it Wavelet Methods for Time Series
Analysis},  Cambridge University Press, Cambridge, UK.

Ram\'irez-Cobo, P. and  Vidakovic, B. (2013). A 2D wavelet-based multiscale
approach with applications to the analysis of digital mammograms.
{\it Computational Statistics \& Data Analysis}, {\bf 58}, 71--81.

Reed, I.S.,   Lee, P.C., and   Truong, T-K. (1995).  Spectral representation of
fractional Brownian motion in n dimensions and its properties. {\it IEEE Transactions on Information
Theory}, {\bf 41}, 5,  1439--1451.

Roberts, T., Newell, M.,  Auffermann, W., and   Vidakovic, B. (2017).
Wavelet-based scaling indices for breast cancer diagnostics. {\it Statistics in
Medicine}, {\bf 36}, 12, 1989--2000.

 Veitch, D. and   Abry, P. (1999). A wavelet-based joint estimator of the parameters of long-range dependence.
{\it IEEE Transactions on Information Theory}, {\bf 45}, 3, 878--897.

 Vidakovic, B. (1999). {\it Statistical Modeling by Wavelets}. John Wiley \& Sons, Hoboken, NJ.

\section{Appendix}
We derive expression (\ref{eq:logE}). For 2D fBf $B_H( x,y) \in \mathbb{R}^2$, detailed wavelet coefficients obtained by NDWT located in one level from the main diagonal hierarchy is
\begin{align}
d_{ (j,j) ; k_1,k_2} &= \int \int B_H({ x,y}) \psi_{j; k_1} ( x)\psi_{j; k_2} ( y) \, dx \, dy \nonumber\\&= 2^{j} \int \int B_H({ x,y})\psi ( 2^{j}(x-k_1))\psi(2^{j}(y-k_2) ) \, dx \, dy,
\label{eq:wavcoef}
\end{align}
where $(j,j) \in \bm j_s$.
We can simplify (\ref{eq:wavcoef}) by letting $\bm k=(k_1,k_2)$ and \textbf v$=(x,y)$.
\ba
d_{(j,j); \bm k} = 2^{j } \int \int B_H(\text{\textbf v})\psi  \big( 2^{j }(\text{\textbf v} - \bm k) \big) \, d\text{\textbf v}.
\ea
The energy of each decomposition level is the variance of the detailed wavelet coefficients $d_{(j,j); \bm k}$ (Heneghan et al., 1996)
\be
E\big[| d_{(j,j); \bm k } |^2\big] & = & 2^{2j } \int \int \psi \big( 2^{j }(\text{\textbf v} - \bm k) \big) \,  \psi \big( 2^{j }(\text{\textbf u} - \bm k) \big) E\big[B_H(\text{\textbf v})B_H(\text{\textbf u})\big] \,d\text{\textbf v} d\text{\textbf u}\nonumber \\
&= &\frac{\sigma^2_H}{2} 2^{2 {j }} \Big[ \int \int \psi \big( 2^{j }(\text{\textbf v} - \bm k) \big) \,  \psi  \big( 2^{j }(\text{\textbf u} - \bm k) \big) |\text{\textbf v}|^{2H} \,d\text{\textbf v} d\text{\textbf u}\nonumber\\ &+&   \int \int \psi  \big( 2^{j }(\text{\textbf v} -\bm k) \big) \,  \psi  \big( 2^{j}(\text{\textbf u} - \bm k) \big) |\text{\textbf u}|^{2H} \,d\text{\textbf v} d\text{\textbf u}
\nonumber\\ & -& \int \int \psi  \big( 2^{j }(\text{\textbf v} - \bm k) \big) \,  \psi  \big( 2^{j }(\text{\textbf u} - \bm k) \big) |\text{\textbf v}-\text{\textbf u}|^{2H} \,d\text{\textbf v} d\text{\textbf u} \Big].
\label{eq:ExpEg}
\ee
By the property of wavelet filters, we know that
\ba
\int \psi  (2^{j}(\text{\textbf v}-\bm k)) d\text{\textbf v}=\int \psi  (2^{j }(\text{\textbf u}-\bm k)) d\text{\textbf u} = 0.
\ea
Thus we can simplify (\ref{eq:ExpEg}) as
\ba
E\big[ |d_{(j,j);\bm k }|^2\big] & = -\frac{\sigma^2_H}{2} 2^{2{j }} \int \int \psi (2^{j }(\text{\textbf v}-\bm k))\psi  (2^{j }(\text{\textbf u}-\bm k)) |\text{\textbf v}-\text{\textbf u}|^{2H}d\text{\textbf v} d\text{\textbf u}.
\ea
We substitute $\bm p=2^{j }(\text{\textbf v}-\text{\textbf u})$ and $\bm q=2^{j }(\text{\textbf u} -\bm k)$ and obtain
\ba
E\big[ |d_{(j,j);\bm k }|^2 \big]&  = &-\frac{\sigma^2_H}{2} 2^{2 j } \int \int \psi(\bm p +\bm q) \psi (\bm q)2^{-2Hj }|\bm p|^{2H} 2^{-4j } d\bm p d\bm q \\
& =& -\frac{\sigma^2_H}{2}2^{-j (2H+2)}\int \int \psi(\bm p + \bm q)\psi (\bm q) |\bm p|^{2H } d\bm p d\bm q\\
& =&\frac{\sigma^2_H}{2}V_{\psi } 2^{-j (2H+2)},
\ea
where $V_{\psi }=-\int \int\psi(\bm p + \bm q)\psi(\bm q) |\bm p|^{2H } d\bm p d\bm q$ which is dependent on wavelet function $\psi$ and Hurst exponent $H$, but independent on $j$. $\sigma_H^2$ is given in (\ref{eq:fbfcov}). By taking logarithm on the energy, we obtain the relationship between wavelet coefficients and the Hurst exponent $H$.
\ba
\log_2 E\Big[ |d_{ (j,j),\bm k}|^2 \Big]=-(2H+2)j +C.
\ea

\end{document}